\documentclass[aps,prl,reprint,superscriptaddress,longbibliography]{revtex4-2}

\usepackage{xcolor}
\usepackage{textcomp}
\usepackage{graphicx} 
\usepackage{float}
\usepackage{epstopdf}
\usepackage{txfonts}

\usepackage[colorlinks,citecolor=blue,linkcolor=blue]{hyperref}

\newcommand{\SPEC}{SPEC, CEA, CNRS, Université Paris-Saclay, Gif-sur-Yvette, France}
\newcommand{\DEE}{Department of Electrical Engineering and ICT, University of Naples Federico II, Naples, Italy} 
\newcommand{\LAF}{Laboratoire Albert Fert, CNRS, Thales, Université Paris-Saclay, Palaiseau, France}
\newcommand{\ITEFI}{Instituto de Tecnologías Físicas y de la Información (CSIC), Madrid, Spain} 
\newcommand{\Zaragoza}{Instituto de Nanociencia y Materiales de Aragón (INMA), CSIC-Universidad de Zaragoza, Zaragoza, Spain}
\newcommand{\Zaragozabis}{Laboratorio de Microscopías Avanzadas (LMA), Universidad de Zaragoza, Zaragoza, Spain}

\begin{document}

\title{Self-modulation instability in high power ferromagnetic resonance of BiYIG nanodisks}

\author{I. Ngouagnia Yemeli}
\affiliation{\SPEC} 
\author{S. Perna}
\email{salvatore.perna@unina.it}
\affiliation{\DEE}
\author{D. Gouéré}
\affiliation{\LAF}
\author{A. Kolli}
\affiliation{\SPEC}
\author{S. Sangiao}
\affiliation{\Zaragoza}
\affiliation{\Zaragozabis}
\author{J. M. De Teresa}
\affiliation{\Zaragoza}
\author{M. Mu\~{n}oz}
\affiliation{\ITEFI}
\author{A. Anane}
\affiliation{\LAF}
\author{M. d'Aquino}
\affiliation{\DEE}
\author{H. Merbouche}
\affiliation{\SPEC}
\author{C. Serpico}
\affiliation{\DEE}
\author{G. de Loubens}
\email{gregoire.deloubens@cea.fr}
\affiliation{\SPEC}

\begin{abstract}

  We study the high power ferromagnetic resonance (FMR) of perpendicularly magnetized BiYIG nanodisks where the uniaxial anisotropy almost compensates the shape anisotropy. We observe a strong saturation of the averaged magnetization upon moderately increasing the amplitude of the rf field and a broadening of the FMR line towards lower and higher magnetic field. Full micromagnetic simulations reveal that a self-modulation of the dynamic magnetization is responsible of this behavior. To get more insight into this unstable dynamics, it is analysed in terms of normal modes. The number of modes involved is found to rapidly increase above the critical threshold. Still, a normal modes model taking into account only a few of them and their mutual nonlinear couplings allows us to qualitatively reproduce the observed phenomenon. The normal modes analysis and micromagnetic simulations also predict a Suhl-like instability at larger excitation power when it is slowly increased from low values, and bistability. Using two-tone spectroscopy, we directly measure the self-modulation spectrum and provide experimental evidence of bistable dynamics. These findings open some perspectives of using high dimensional dynamics in magnetic nanostructures for unconventional information processing.

\end{abstract}

\maketitle

Nonlinear magnetization dynamics \cite{wigen94,lvov94,gurevich96,mayergoyz09} exhibit various phenomena, which have been extensively studied for several decades. They range from spin-wave (SW) instabilities \cite{anderson55,suhl57,bertotti01a,loubens05,gerrits07}, auto-oscillations and chaos \cite{gibson84,rezende90,petit-watelot12,ustinov21}, to magnetic solitons \cite{kosevich90,wu06,mohseni13} and the condensation of magnons \cite{demokritov06,schneider20,divinskiy21a}. Beyond the fundamental interest of such phenomena and their counterparts in other fields of physics, perspectives of using nonlinear magnetization dynamics for unconventional information processing with magnonic and spintronic devices have recently emerged, such as for symbolic dynamics \cite{yoo20}, random number generation \cite{phan24}, and neuromorphic computing \cite{koerber23}.

Ferromagnetic resonance (FMR) is well adapted to investigate nonlinear phenomena. There, a uniform rf field is used to resonantly drive the dynamical system in the strong out-of-equilibrium regime \cite{bertotti01}. In large ferromagnetic samples, such as millimeter sized yttrium iron garnet (YIG) spheres, a rich phase diagram emerges due to interactions between quasi degenerate SW modes \cite{bryant88}. In out-of-plane magnetized thin films, the main FMR mode lies at the bottom of the SW spectrum, and by decreasing the sample dimensions, the density of normal modes can be lowered, with signatures of this discretization in the nonlinear regime \cite{mcmichael90,naletov07}. In nanostructures, the strong geometrical confinement completely lifts the degeneracy between quantized SW modes, so that energy and momentum conservation laws are more difficult to fulfill for nonlinear scattering processes \cite{melkov13}. As a result, SW instability thresholds are pushed back to higher power, and spatially coherent magnetization precession can be driven at very large amplitude in submicron disks, accompanied by a strong shift of the resonance line and bistable foldover due to the large nonlinear frequency shift \cite{li19c}. 

This nonlinear frequency shift can be controlled by adjusting the total effective anisotropy of the thin film, in particular, its sign can be tuned \cite{luehrmann91}. In the situation where it is close to zero, large angles of coherent precession can be achieved without significant shift of the resonance conditions \cite{gnatzig87}, which can be useful for the realization of certain phenomena. In fact, spin-orbit torque induced emission of coherent propagating SWs \cite{evelt18a}, SW amplification \cite{merbouche24} and magnon condensation \cite{divinskiy21a} were recently demonstrated based on hybrid BiYIG / platinum bilayers, which was only possible thanks to the uniaxial anisotropy compensating the shape anisotropy in the thin bismuth doped YIG layer \cite{soumah18,gouere22}.

In this context, we investigate the nonlinear FMR of out-of-plane magnetized BiYIG submicron disks with nearly vanishing effective anisotropy. From the above arguments, one would expect that large amplitude single mode oscillation without nonlinear frequency shift could be achieved. Instead, we observe a strong saturation of the FMR line when the amplitude of the rf field is increased, together with the appearance of a frequency modulated spectrum and of bistable dynamics. Based on micromagnetic simulations and a normal modes model, we ascribe these surprising results to the onset of self-modulation of the precession profile, where dipolar interactions between quantized SW modes play a major role. Our analysis shows that above a threshold power there is a chaotic attractor that is created, which, for certain microwave power, coexists with an oscillation synchronous to the rf field. For higher rf field power, there is a Suhl-like mechanism where such oscillation becomes unstable and chaotic motion is the only possible observable. Moreover, the number of modes involved in the chaotic dynamics depends exponentially on the rf field amplitude. These phenomena affect the FMR response which looks very different from the typical one of a uniform magnet with uniaxial anisotropy.

The studied sample is composed of a few BiYIG disks with diameters 500, 700 and 1000~nm patterned from a 30~nm Bi$_1$Y$_2$Fe$_5$O$_{12}$ thin film grown by pulsed laser deposition on a (111)-oriented substituted-GGG substrate, optimized to exhibit close to zero effective magnetization, $\mu_0 M_{\text{eff}} = -10$~mT, and low damping, $\alpha = 7 \cdot 10^{-4}$ \cite{gouere22}. A 4.5~$\mu$m wide gold microwave antenna is patterned on top of them to supply a spatially uniform, linearly polarized in-plane rf field. The BiYIG disks are saturated by a magnetic field $H_z$ applied along the normal $z$ to the plane. We use a magnetic resonance force microscope (MRFM) to perform the FMR spectroscopy of the individual disks \cite{klein08}. The MRFM probe, made of a cobalt nanosphere grown at the apex of a soft cantilever \cite{sangiao17}, sensitively monitors the averaged longitudinal component of magnetization $M_z$ of the BiYIG disk under rf excitation, which is pulse modulated at the cantilever frequency, $f_c\simeq 13$~kHz.

\begin{figure}
	\centering
        \includegraphics[width=8cm]{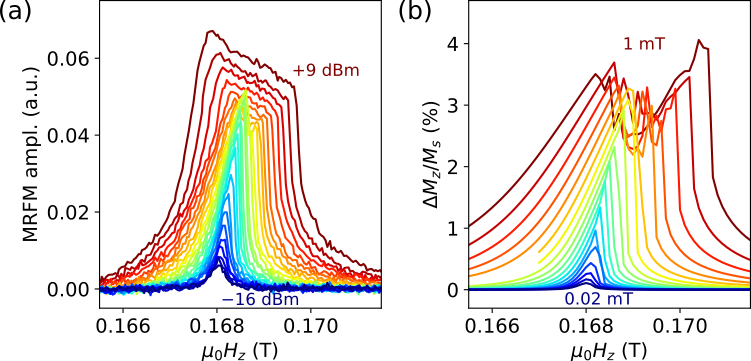}
	\caption{(a) Out-of-plane MRFM spectroscopy at 5 GHz of a BiYIG nanodisk of diameter 700~nm as a function of rf power, varied from $-16$~dBm to $+9$~dBm by steps of 1 dBm. (b) Longitudinal reduction of magnetization extracted from micromagnetic simulations as a function of rf field amplitude, varied from 0.02~mT to 1~mT by logarithmic steps.}
	\label{fig:expsim}
\end{figure}

The broadband MRFM spectroscopy performed in the linear regime confirms the good dynamical quality of the BiYIG disks, which all have a damping close to $10^{-3}$. In the following, we fixed the frequency of the rf field to 5 GHz. Figure \ref{fig:expsim}(a) presents the evolution of the main FMR line of the 700~nm nanodisk, as a function of the rf power $P$, varied from $-16$~dBm to $+9$~dBm. At low $P$, it has a lorentzian shape with a linewidth of 0.3~mT. When $P$ is increased, the FMR line first grows in amplitude and slightly shifts toward higher field, which is expected from the small and negative effective magnetization of the BiYIG. However, we observe a qualitative change of its aspect beyond $P \approx 0$~dBm (green-yellow spectra in Fig.\ref{fig:expsim}(a)): it strongly saturates in amplitude and broadens in both high and low field directions. A similar behavior is observed for all BiYIG disks with diameters ranging from 500 to 1000~nm \cite{supplemental}.

To understand these experimental results, we performed micromagnetic simulations \cite{magnum} using magnetic parameters extracted from linear MRFM spectroscopy. The 700~nm diameter disk was discretized using a 64$\times$64$\times$1 rectangular mesh. Figure \ref{fig:expsim}(b) shows the evolution of $\Delta M_z/M_s = 1 - M_z/M_s = 1 - m_z$ versus $H_z$, swept downwards by steps of 0.1 mT, for different rf field amplitudes spanning from 0.02~mT to 1~mT. For each $H_z$, the spatially averaged cartesian components of the magnetization are recorded for 100~ns under an harmonic excitation at 5~GHz of constant amplitude $h_{\text{rf}}$. The plotted quantity is obtained by averaging $m_z$ over the second half of the simulated time window. A good qualitative agreement with the MRFM spectroscopy data is observed. In particular, the breakdown of the FMR line is reproduced beyond $\mu_0h_{\text{rf}}\approx 0.4$~mT, and $\Delta M_z/M_s$ saturates around 3 to 4\%. 

\begin{figure}
	\centering
        \includegraphics[width=8.5cm]{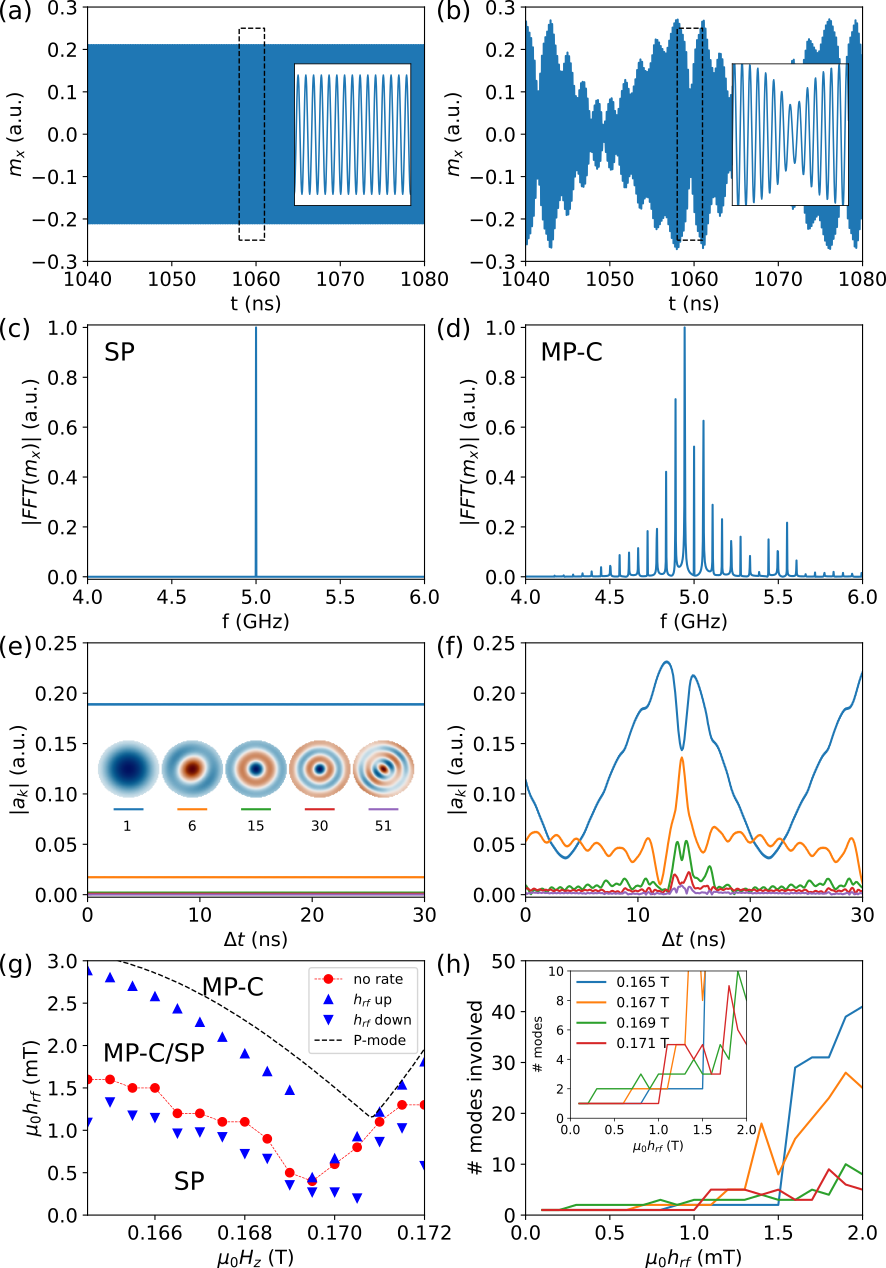}
	\caption{Micromagnetic simulations. (a,b) Time series of $m_x$, (c,d) corresponding power spectra and (e,f) mode amplitudes vs. time at $\mu_0H_z = 0.169$ T, respectively below ($\mu_0h_{\text{rf}} = 0.4$ mT) and beyond ($\mu_0h_{\text{rf}} = 0.6$ mT) the critical threshold. (g) Phase diagram in the control plane ($H_z$, $h_{\text{rf}}$). $\mu_0H_z$ is varied from 0.1645 to 0.172~T by steps of 0.5 mT and $\mu_0h_{\text{rf}}$ from 0.1 to 3~mT by steps of 0.1~mT, the rf frequency being fixed to 5~GHz. Red dots correspond to the critical curve obtained by directly applying the rf field to the equilibrium state, blue upwards (downwards) triangles to the one obtained by increasing (decreasing) slowly $h_{\text{rf}}$ from zero (maximum) amplitude. The dashed curve is the instability threshold deduced from \textbf{P}-mode analysis. (h) Dependence on $h_{\text{rf}}$ of the number of modes with amplitudes $|a_k| > 10^{-3}$ involved in the dynamics at four different $H_z$.
          }
	\label{fig:sim}
\end{figure}

To determine the phase diagram of our dynamical system in the control plane ($H_z$, $h_{\text{rf}}$), we performed additional simulations using the MaGICo micromagnetic code \cite{magico}. For each $H_z$ value, the equilibrium configuration is excited with an rf field of frequency 5~GHz and amplitude $h_{\text{rf}}$ and the dynamics is integrated for 1100~ns. The last 100 ns are used to evaluate the type of regime. The time series of the transverse magnetization $m_x = M_x/M_s$ are Fourier transformed to obtain their power spectrum. At low $h_{\text{rf}}$, $m_x$ oscillates at constant amplitude at the forced frequency of 5~GHz, see Fig.\ref{fig:sim}(a), resulting in a single-peak (SP) spectrum such as shown in Fig.\ref{fig:sim}(c). At sufficiently high power, however, it does not display a time-harmonic response anymore. Instead, one observes a low frequency modulation of its amplitude, see Fig.\ref{fig:sim}(b), and a multi-peaks continuous (MP-C) spectrum such as in Fig.\ref{fig:sim}(d). The rf field amplitude responsible for the transition between SP and MP-C dynamical states is plotted as a function of $H_z$ using red dots in Fig.\ref{fig:sim}(g), which defines a critical curve. Simulations have also been run with $h_{\text{rf}}$ being slowly varied. First, $h_{\text{rf}}$ is increased starting from a SP state until the oscillations become MP-C (upwards triangles). Then, $h_{\text{rf}}$ is decreased starting from a MP-C state until the oscillations become SP (downwards triangles). Interestingly, there is a region in the control plane ($H_z$, $h_{\text{rf}}$) where there is coexistence between oscillations with MP-C and SP spectra. We also observe that the critical curve estimated by directly applying $h_{\text{rf}}$ (red dots) quantitatively agrees with the curve obtained when $h_{\text{rf}}$ is slowly decreased from a MP-C state (downward triangles).

To analyse in more details the underlying dynamics, we project snapshots of the magnetization field from micromagnetic simulations onto the basis of normal modes \cite{massouras24}. The latter are first obtained numerically by linearizing the Landau-Lifshitz (LL) equation, which yields a generalized eigenvalue problem that can be solved \cite{aquino09}, and ordered by ascending frequency \cite{supplemental}. The fundamental mode is the uniform or Kittel mode. Other normal modes are quantized modes along the azimuthal and radial directions. Only the purely radial modes couple to the uniform rf field excitation \cite{naletov11}. The profiles of five of them, 1, 6, 15, 30 and 51, are shown as insets in Fig.\ref{fig:sim}(e). The mode amplitudes evaluated in this way are plotted as a function of time in Figs.\ref{fig:sim}(e,f). Before the threshold, only modes 1 and 6 have significant amplitude. The one of the uniform mode, excited close to resonance by the rf field, is by far the largest and is constant in the steady state. Above the threshold, it strongly oscillates and the amplitudes of other radial modes start to be significant and oscillating, as shown in Fig.\ref{fig:sim}(f). This oscillation of modes is responsible for the one of the precession profile visualized in movies extracted from micromagnetic simulations, in which a strong spatio-temporal modulation of $m_z$ in the disk is observed \cite{supplemental}.

We can also estimate how many modes are involved in the dynamics from the micromagnetic simulations. Fig.\ref{fig:sim}(h) displays the number of normal modes with significant amplitude ($|a_k| > 10^{-3}$) as a function of $h_{\text{rf}}$ for several values of $H_z$. Interestingly there are two types of MP-C states: one where the number of modes involved is around 10 and the other where it becomes several times that. When $\mu_0H_z= 0.169$~T, it is less than 5 until $\mu_0h_{\text{rf}}$ reaches 1.9~mT, where it increases to 10. A more dramatic behavior is observed at $\mu_0H_z= 0.165$~T, where the number of modes involved explodes from 2 to 30 above 1.5~mT. 

\begin{figure}
	\centering
        \includegraphics[width=8.5cm]{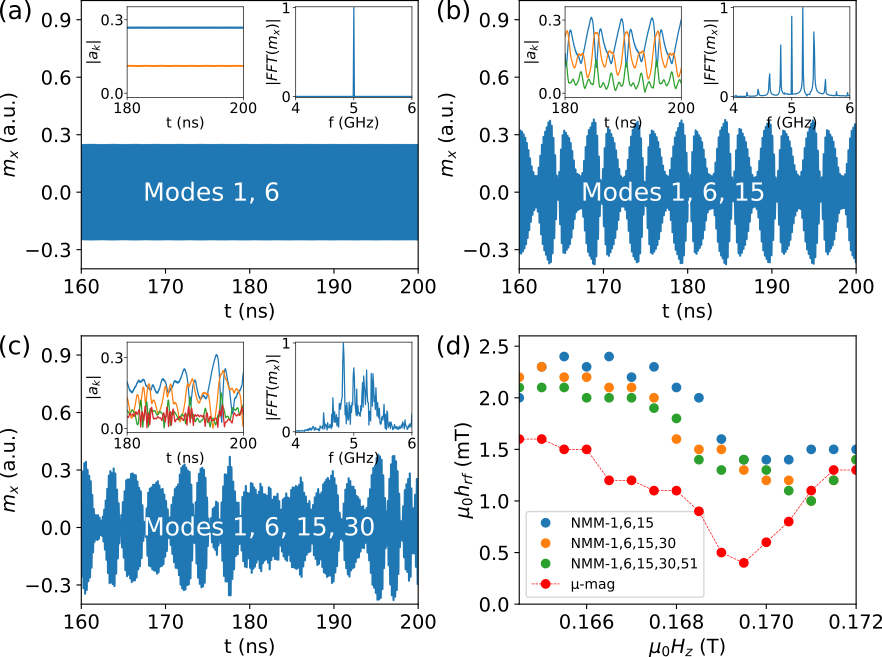}
	\caption{Time series of $m_x$ (insets: mode amplitudes vs. time and power spectra) calculated at $\mu_0H_z = 0.172$~T and $h_{\text{rf}} = 3$~mT using the NMM including (a) two (b) three and (c) four modes. (d) Phase diagram in the control plane ($H_z$, $h_{\text{rf}}$) comparing the critical curve extracted from full micromagnetics (red dots) to the ones using the NMM with three, four and five modes (blue, orange and green dots, respectively).
        }
	\label{fig:NMM}
\end{figure}

Micromagnetic simulations show that several SW modes participate to the dynamics in the nonlinear regime. Hence, a normal modes model (NMM) taking into account the relevant modes should be appropriate to reproduce the self-modulation instability. Because it only considers a few degrees of freedom, it is computationally efficient and it can bring useful physical insight. Shortly, the LL equation is projected onto the normal modes basis, resulting in a system of nonlinear ordinary differential equations for the mode amplitudes, whose coefficients can be pre-computed from the normal modes determined previously \cite{perna22,perna22a}. We first consider a NMM with only modes 1 and 6, which have the largest amplitudes in micromagnetics. Integrating the two coupled equations yields a stable dynamics at $\mu_0h_{\text{rf}}=3$~mT, see Fig.\ref{fig:NMM}(a); it becomes unstable only at very large amplitude ($> 4$~mT). By adding mode 15 into the NMM, the self-modulation is already present at 3~mT, see Fig.\ref{fig:NMM}(b). As can be seen in Fig.\ref{fig:NMM}(c), it remains upon including in the dynamics mode 30, although the spectrum changes. Using the NMM with three modes, one can construct the critical line in the ($H_z$, $h_{\text{rf}}$) plane when the rf field is pulsed from equilibrium and compare it to the one obtained from full micromagnetic simulations. A qualitative agreement is observed in Fig.\ref{fig:NMM}(d), but the threshold is overestimated by the NMM. The inclusion of more modes improves the situation, still, quantitative agreement could not be obtained with only five modes. Integrating the NMM when $h_{\text{rf}}$ is slowly varied also shows the presence of a coexistence region between SP and MP-C states \cite{supplemental}. The curve for increasing $h_{\text{rf}}$ is quite close to the one computed micromagnetically, however, the curves for decreasing $h_{\text{rf}}$ differ quantitatively. This is ascribed to the fact that in the MP-C dynamical state the number of modes involved is higher than that considered in the NMM.

\begin{figure}
	\centering
        \includegraphics[width=8cm]{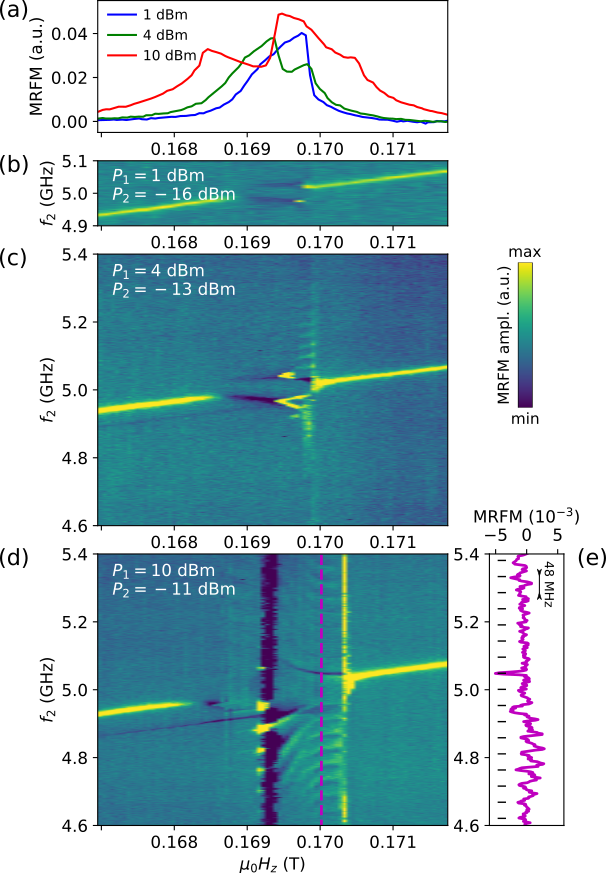}
	\caption{(a) Single-tone MRFM spectroscopy at 5~GHz of a BiYIG nanodisk of diameter 500~nm for three different rf powers (pulse modulated). (b), (c) and (d) Two-tone MRFM spectroscopy as a function of the up-swept field $H_z$ and the frequency $f_2$ of the low power rf excitation (pulse modulated). The power of the continuous rf excitation at $f_1=5$~GHz is fixed respectively to 1, 4 and 10~dBm. (f) MRFM spectrum as a function of $f_2$ extracted at fixed $H_z$ from the spectroscopy data, indicated by the vertical dashed line in (e).}
	\label{fig:2tone}
\end{figure}

We also performed a \textbf{P}-mode analysis, where the amplitudes of normal modes are calculated at equilibrium in a rotating frame synchronous with the rf field angular frequency $\omega$ and the \textbf{P}-modes equations are derived from the hypothesis $a_k=\overline{a_k}e^{j\omega t}$, which allows to define a Suhl-like instability threshold \cite{supplemental}. It is found to be in good agreement with the threshold estimated from simulations when slowly increasing $h_{\text{rf}}$, see dashed line and upwards triangles in Fig.\ref{fig:sim}(g).

To provide some experimental evidence of the scenario deduced from micromagnetics and normal modes analysis, we performed two-tone MRFM spectroscopy \cite{li19c}. In these measurements, a continuous wave rf excitation of fixed frequency ($f_1=5$~GHz) and high power $P_1$ is applied to the 500~nm BiYIG disk to drive its dynamics in the nonlinear regime, while a second rf tone of low power $P_2$ and varying frequency $f_2$ is pulse modulated at the cantilever frequency to probe the dynamical state driven by the first rf tone. At $P_1=1$~dBm, the single-tone MRFM spectroscopy (blue spectrum in Fig.\ref{fig:2tone}(a)) shows no sign of instability, and the two-tone spectroscopy map of Fig.\ref{fig:2tone}(b) is a signature of a coherent \textbf{P}-mode dynamics, with two symmetric peaks around $f_1$ reflecting the nutation of the magnetization vector around its large amplitude steady state trajectory \cite{li19c}. At $P_1=4$~dBm, the single-tone MRFM spectrum plotted in green in panel (a) strongly distorts and saturates for $0.1695<\mu_0H_z<0.17$~T, and in this region, the two-tone spectroscopy displays a bifurcation of the two nutation peaks followed by a broad multi-peaks spectrum (Fig.\ref{fig:2tone}(c)), which reflects the transition to a modulation instability. Finally at $P_1=10$~dBm, the instability threshold is crossed for a broad range of applied field (red spectrum in panel (a), $0.1685<\mu_0H_z<0.1705$~T). As a consequence, the two-tone spectroscopy map is multi-peaks almost everywhere, see Fig.\ref{fig:2tone}(d). A $f_2$-cut of the data at $\mu_0H_z=0.17$~T displaying a modulation frequency of 48~MHz, in the same range as found in simulations, is shown in Fig.\ref{fig:2tone}(d).

\begin{figure}
	\centering
        \includegraphics[width=8.5cm]{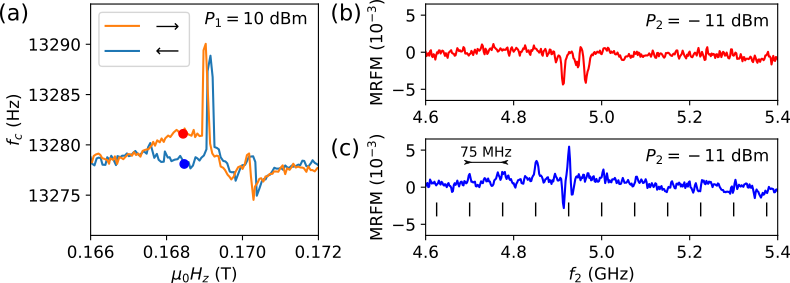}
	\caption{(a) MRFM cantilever frequency as a function of $H_z$ under continuous rf excitation at $f_1=5$~GHz. (b) and (c) Two-tone MRFM spectroscopy performed at $\mu_0H_z=0.1685$~T, coming respectively from low and high field (see red and blue dots in (a)).}
	\label{fig:bistab}
\end{figure}

To experimentally access the bistability between SP and MP-C dynamical states, we monitor the frequency of the MRFM cantilever under continuous rf excitation when $H_z$ is swept up and down through the FMR line, as shown in Fig.\ref{fig:bistab}(a). An hysteretic behavior is observed between 0.1675 and 0.169~T. The two-tone MRFM spectroscopy performed at $\mu_0H_z=0.1685$~T coming from low and high field, shown in Fig.\ref{fig:bistab}(b) and (c), displays distinct signatures, assigned to coherent and frequency modulated dynamics, respectively \cite{supplemental}.

Finally, we briefly discuss the origin of the dynamic instability observed when driving these BiYIG nanodisks into the nonlinear regime of FMR. Due to finite size effects and the non-ellipsoidal disk geometry, the fundamental mode excited by the uniform rf field has a non-uniform amplitude, despite having a uniform precession phase (mode 1 in Fig.\ref{fig:sim}(e)). When its amplitude grows, it usually cannot maintain its profile, instead it spatially extends or contracts depending on the sign of the nonlinear frequency shift \cite{gulyaev00,slavin05c,demidov10,li19c}. When the latter is close to zero, it can keep its non-uniform profile in the nonlinear regime. The strong non-uniform dipolar field that it generates in the magnetic body makes it possible to excite other SW modes, which couple weakly to the uniform rf excitation, at a substantial level (particularly mode 6, see Fig.\ref{fig:sim}(e,f)). There is thus an energy from long to shorter wavelength modes, which can lead to the modulation instability described above. 

To conclude, we would like to emphasize that two types of phenomena have been unveiled by our theoretical analysis: the coexistence of oscillation synchronous with rf field (SP) and chaotic attractor (MP-C) at moderate power, and a Suhl-like instability at higher power, where the SP state becomes unstable. We have also found that the number of modes involved in the chaotic dynamics depends exponentially on power, which limits the accuracy of the normal modes model, but at the same time opens opportunities to study turbulence in confined systems. Moreover, detailed analysis of this chaotic regime is under investigation, as well as the possibility to use chaotic modes motion and multi-stability of dynamical regimes to perform computations.

This work was supported by the Agence Nationale de la Recherche (ANR) under grant no. ANR-18-CE24-0021 (Maestro), ANR-20-CE24-0012 (Marin), the PEPR SPIN project ANR-22-EXSP-0005 (SpinCom), and by the Horizon2020 Research Framework Programme of the European Commission under grant no. 899646 (k-NET). It was also supported by a public grant overseen by the ANR as part of the ``Investissements d'Avenir'' program (Labex NanoSaclay, reference: ANR-10-LABX-0035). S.S. and J.M.D.T thank the Gobierno de Aragón through the grant numbers E13\_23R and Ayuda CEX2023-001286-S funded by MICIU/AEI/10.13039/501100011033 and European Union “NextGenerationEU”/PRTR.

\end{document}


\title{Supplementary Material: ``Self-modulation instability in high power ferromagnetic resonance of BiYIG nanodisks''}

\author{I. Ngouagnia Yemeli}
\affiliation{\SPEC} 
\author{S. Perna}
\email{salvatore.perna@unina.it}
\affiliation{\DEE}
\author{D. Gouéré}
\affiliation{\LAF}
\author{A. Kolli}
\affiliation{\SPEC}
\author{S. Sangiao}
\affiliation{\Zaragoza}
\affiliation{\Zaragozabis}
\author{J. M. De Teresa}
\affiliation{\Zaragoza}
\author{M. Mu\~{n}oz}
\affiliation{\ITEFI}
\author{A. Anane}
\affiliation{\LAF}
\author{M. d'Aquino}
\affiliation{\DEE}
\author{H. Merbouche}
\affiliation{\SPEC}
\author{C. Serpico}
\affiliation{\DEE}
\author{G. de Loubens}
\email{gregoire.deloubens@cea.fr}
\affiliation{\SPEC}

\maketitle

\tableofcontents

\clearpage
\newpage


\section{MRFM spectroscopy vs. disk diameter}

\begin{figure}
	\centering
        \includegraphics[width=16cm]{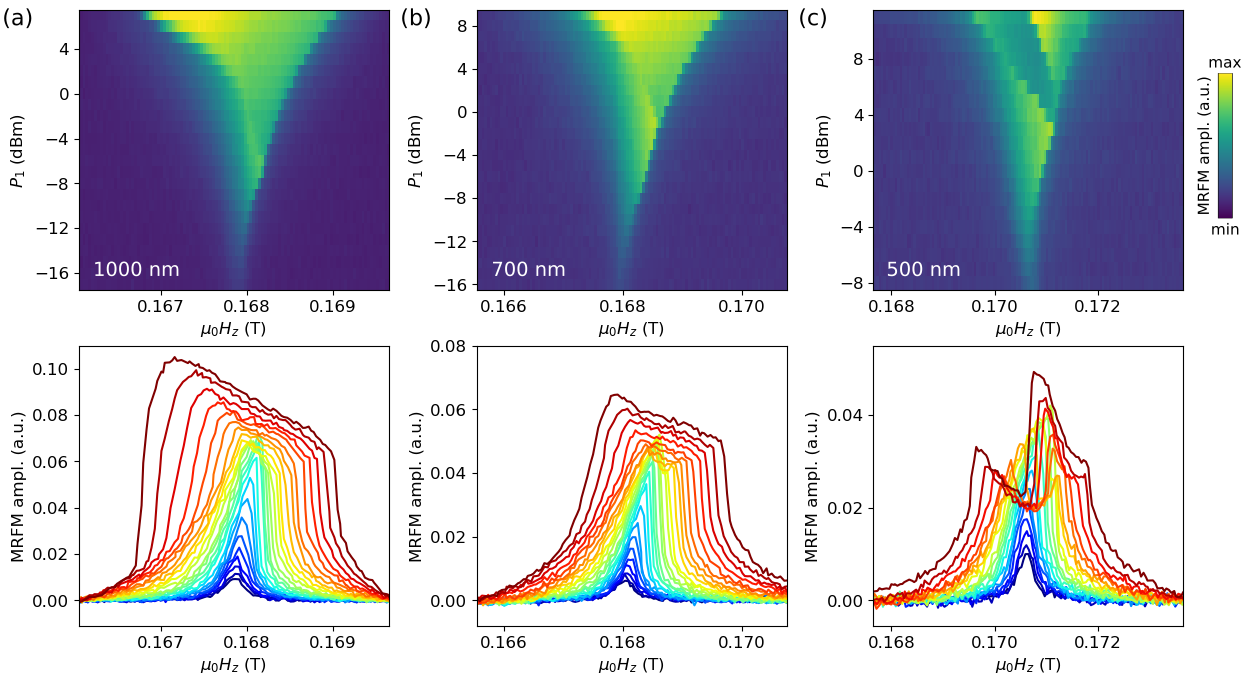}
	\caption{Dependence on disk diameter. MRFM spectroscopy performed at 5~GHz as a function of the excitation power for BiYIG nanodisks of diameters 1000~nm (a), 700~nm (b), and 500~nm (c). The data plotted in (b) are the same as in Fig.1(a) of the main text.}
	\label{fig:expvsD}
\end{figure}

For each diameter among 500, 700 and 1000~nm, there are two BiYIG disks available under the microwave antenna. We performed MRFM spectroscopy on all of them, and observed a similar saturation and broadening of the main FMR line as the rf amplitude is increased. Suppl. Fig.\ref{fig:expvsD} shows these measurements for one particular disk of each diameter. Upper and lower panels show the same data plotted as 2D colormaps and 1D spectra vs. power, respectively.

\section{Supplementary movies}

The supplementary movies show the dynamics calculated using micromagnetic simulations at $\mu_0H_z=0.169$~T and $\mu_0h_{\text{rf}}=1$~mT (rf frequency equal to 5~GHz) in the 700~nm disk as a function of time. The total simulated time is equal to 130~ns, the last 30~ns being displayed in the movies.

The upper panel of SupMov1.gif shows the spatial average over the disk volume of the magnetization components along $x$, $y$ and $z$, in blue, green and red lines, respectively. The transverse magnetization, defined as $m_t=\sqrt{m_x^2+m_y^2}$ is also shown in cyan. The components $m_x$ and $m_y$ oscillate in quadrature at 5~GHz, and exhibit an amplitude modulation of period around 10~ns (i.e., a modulation frequency of about 100~MHz), which can also be clearly seen on the $m_z$ component. The lower left and right panels show the evolution as a function of time of the 1D spatial profiles of the transverse ($m_t$) and longitudinal ($m_z$) components vs. the reduced radial coordinate $r/R$, respectively ($R=350$~nm is the disk radius). One clearly observes spatio-temporal oscillations of the $m_t$ spatial profile, accompanied by a strong modulation of $m_z$, which exhibits large and sudden reductions at the disk center. There, $m_z$ can even become negative at $t \approx 12$~ns, \textit{i.e.}, the angle of precession $\theta$ becomes larger than $\pi/2$ ($m_z=-0.5$ corresponds to $\theta = 2\pi/3$).

The upper panel of SupMov1bis.gif is the same as SupMov1.gif. The three lower panels display, from left to right, the time evolution of the 2D spatial profiles of the precession phase, $\phi = \arctan{(m_y/m_x)}$, of the transverse component, $m_t$, and of the longitudinal one, $m_z$. In the lower left panel, the hue color scale codes the $[0,2\pi]$ interval, while in the other two panels, the gray contrast codes the $[0,1]$ interval (white: 0, black:1). One can observe that the phase is not constantly uniform in the disk, instead it shows periodic modulation of its spatial profile, as do the $m_t$ and $m_z$ components. 


\section{Normal modes model}

\subsection{~Calculation of normal modes}

\begin{figure}
	\centering
        \includegraphics[width=12cm]{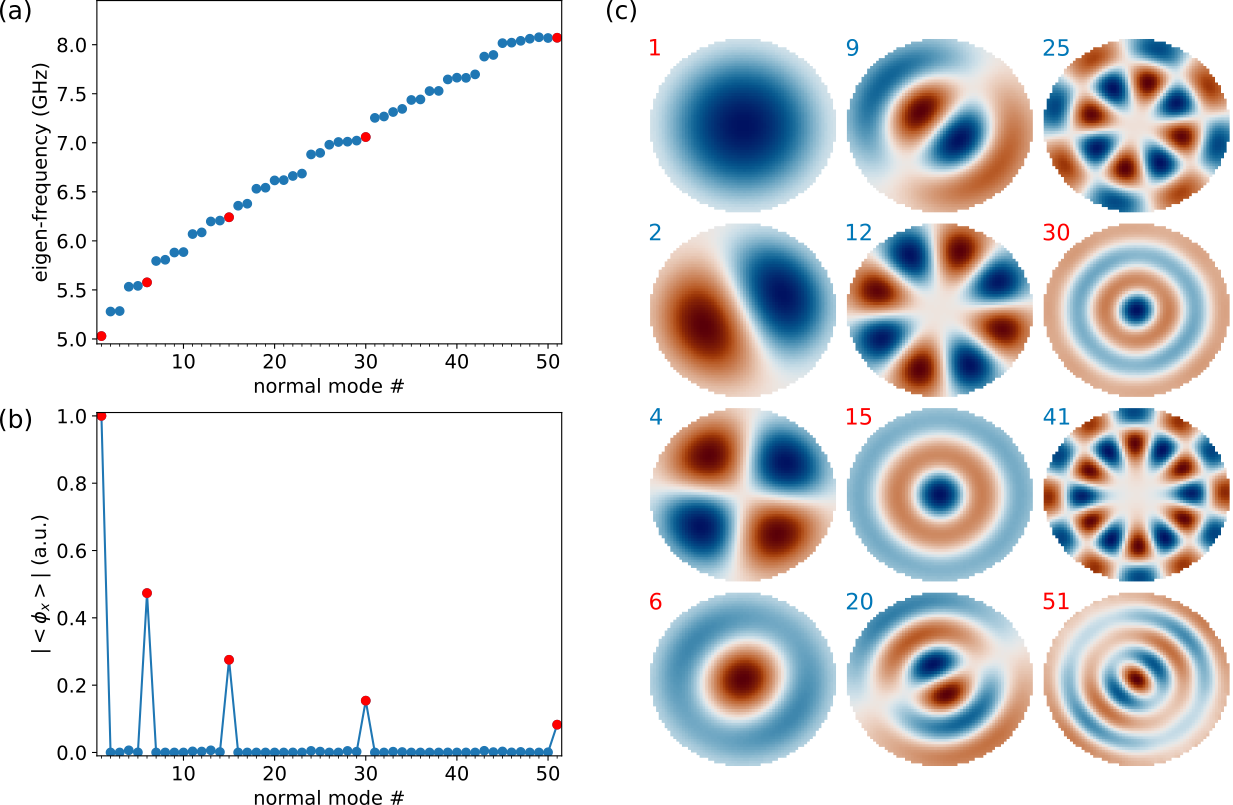}
	\caption{Micromagnetic normal modes. (a) Normal mode spectrum of the 700 nm BiYIG disk calculated at $\mu_0H_z=0.169$~T. (b) Absolute value of the spatial average of the x-component of normal modes vs. mode number. (c) Mode profiles of 12 chosen normal modes. The red/blue color represents the phase of the precession, while the contrast represents its amplitude. Red labels emphasize radial modes 1, 6, 15, 30, 51 having a nonzero average in (b).}
	\label{fig:magico}
\end{figure}

The normal modes are calculated by MaGICo \cite{magico} which solves the generalized eigenvalue problem for normal modes according to the formulation presented in \cite{aquino09}. Suppl. Fig.\ref{fig:magico} displays the normal modes spectrum (a), the absolute value of the spatial average of the x-component (b) of normal modes, and the norm of the real part of normal modes (c). Among the first 60 normal modes, 5 of them have significative average along the x-direction: 1, 6, 15, 30, 51, see Suppl. Fig.\ref{fig:magico}(b). Their respective frequencies are outlined with red filled dots in Suppl. Fig.\ref{fig:magico}(a). They correspond to radial modes with azimuthal invariance, as shown in Suppl. Fig.\ref{fig:magico}(c).


\subsection{~Projection of LLG equation on normal modes basis}

When the micromagnetic equilibrium is slightly perturbed, the magnetization dynamics can be decomposed in harmonic and independent oscillations of normal modes at their respective natural frequency. The set of normal modes can also be used as a basis to project the nonlinear magnetization dynamics described by the LLG equation. In this way, the following system of coupled nonlinear equations is derived where the unknowns are the amplitudes of normal modes:
\begin{equation}\label{eq:NMM}
  \dot{a_k} = \frac{j\omega_k}{(1+\alpha^2)} \left[ b_k + \sum_{h} b_{k,h}a_h + \sum_{h,i} c_{k,h,i}a_ha_i + \frac{1}{2} \sum_{h,i,j} d_{k,h,i,j}a_ha_ia_j + ... \right]\,.
\end{equation}
This is the so-called normal modes model (NMM) \cite{perna22}. In several cases of interest for applications, only few modes and few terms in the NMM are required for a quantitative analysis of the micromagnetic system under investigation. In this way, a hierarchy of approximated models can be defined, whose description is intermediate between the macrospin (uniform mode) approximation and the full micromagnetic description.


\subsection{~Reduced normal modes model}

Here, we consider normal modes that have a significative spatial average along the rf field direction (x). Such modes correspond to index numbers 1, 6, 15, 30, 51,... We take only the first 5 of them, see Suppl. Fig.\ref{fig:magico}. Not all the terms of the NMM are considered in the numerical analysis with normal modes. Indeed, the spectrum of the averaged magnetization components is centered around the rf field frequency of 5~GHz. The NMM is then written in a rotating frame at the angular frequency of the rf field, $\omega_{\text{rf}}$, where the k-th amplitude of normal mode is expressed as $a_k(t) = \bar{a}_k(t) e^{j\omega_{\text{rf}}t}$ and $\bar{a}_k(t)$ is a slow time-varying function.
In this way, we get the following reduced NMM:
\begin{equation}\label{eq:NMM_RWA}
  \dot{\bar{a}_k} = \frac{j\omega_k}{(1+\alpha^2)} \left[ \bar{b}_k + \left(b_{k,k}-\frac{\omega_{\text{rf}}}{\omega_k}\right)\,\bar{a}_k + \sum_{h,i} \left(c_{k,h,-i}^{f,cw}\bar{a}_h\bar{a}_i^*+c_{k,-h,i}^{f,cw}\bar{a}_h^*\bar{a}_i+c_{k,h,i}^{f,ccw}\bar{a}_h\bar{a}_i\right) 
  + \frac{1}{2} \sum_{h,i,j} \left(d_{k,h,i,-j}^0\bar{a}_h\bar{a}_i\bar{a}_j^* + d_{k,h,-i,j}^0\bar{a}_h\bar{a}_i^*\bar{a}_j+d_{k,-h,i,j}^0\bar{a}_h^*\bar{a}_i\bar{a}_j\right) \right]\,,
\end{equation}
where all the terms $b_{k,h}$ with $h\neq k$ have been neglected. The terms $c_{k,-h,i}^{f,cw}$ are the coefficients of the three-magnon scattering that are due to the rf field. The superscript cw means that the clockwise rotating component of the rf field is considered for its evaluation. In the same way, ccw stands for counter-clockwise (the rf field being linearly polarized, it can be decomposed into its cw and ccw components). The terms $d_{k,h,i,j}^0$ are the coefficients of the four-magnon scattering. The zero superscript indicates that they are derived from conservative torque terms of the LLG. The other terms derived from non-conservative torques have been neglected because of their relatively small value.


\subsection{~\textbf{P}-mode analysis}

\begin{figure}
	\centering
        \includegraphics[width=12cm]{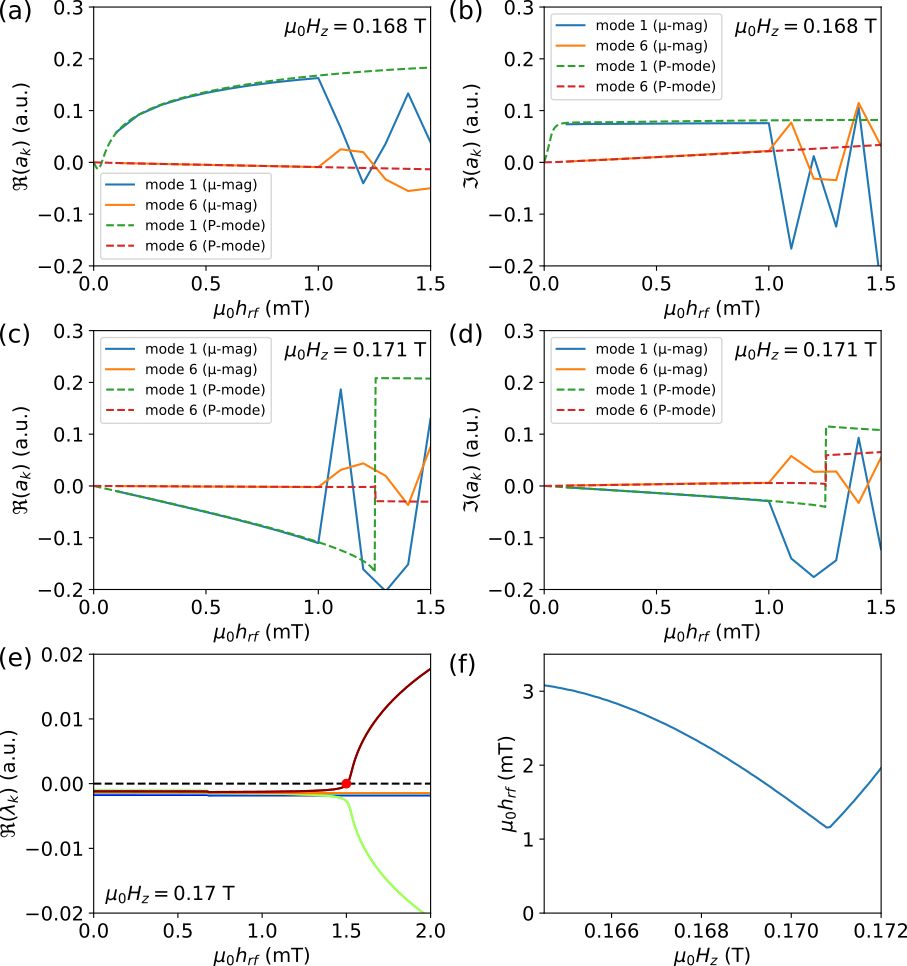}
	\caption{\textbf{P}-mode analysis. Comparison of the real and imaginary parts of modes 1 and 6 extracted vs. rf field amplitude from micromagnetic simulations (continuous lines) and \textbf{P}-mode analysis (dashed lines) at (a), (b) $\mu_0H_z=0.168$~T and (c), (d) $\mu_0H_z=0.171$ ~. (e) Dependence on $h_{\text{rf}}$ of the real part of \textbf{P}-mode eigenvalues at $\mu_0H_z=0.17$~T. (f) Instability threshold vs. $H_z$ from \textbf{P}-mode analysis.}
	\label{fig:P-modes}
\end{figure}

For small rf power, the driven magnetization oscillations in micromagnetic simulations are time harmonic. The spectrum of such oscillations shows a single peak (SP) at the same frequency of the driving field, which therefore will be termed SP solution. Then, we look for solutions of eq.\ref{eq:NMM_RWA} of the following type $a_k(t) = \bar{a}_k e^{j\omega_{\text{rf}}t}$, where this time $\bar{a}_k$ is a phasor \emph{not} depending on time. In this way, the reduced NMM becomes a set of algebraic nonlinear equations whose solution gives the phasors of the normal modes. Such a set will be called in the following \textbf{P}-mode equation. 

For a fixed initial solution and fields values $(H_z,h_{\text{rf}})$, the solutions of the \textbf{P}-mode equation are computed using the Newton-Raphson method. Starting from zero rf field where all modes have zero amplitude, numerical continuation is used and a sequence of solutions $(h_{\text{rf},n},\{\bar{a}_1,\bar{a}_2,\dots\}_n)$ for $n = 0,1,\dots$ is calculated at fixed $H_z$. These solutions, which in time domain represent modes with time harmonic oscillations and fixed phases, are then compared with the amplitudes of normal modes computed by full micromagnetic simulations under the same excitation conditions. For each couple $(H_z,{h_{\text{rf},n}})$, taking the equilibrium as initial condition, a simulation of $1100$~ns is performed and the snapshot of the final magnetization field is saved. The duration of the simulation is such that the SP solution is reached for values of $h_{\text{rf}}$ where it exists. These snapshots are then projected onto the basis of normal modes, so that their complex amplitudes can be computed. The projection technique is discussed in the section Methods of ref.\cite{perna22}. The comparison of the real and imaginary parts of normal modes computed using the \textbf{P}-mode equation and micromagnetic simulations is shown as a function of $h_{\text{rf}}$ for two values of $H_z$ in Fig.\ref{fig:P-modes}(a)-(d). Starting from $h_{\text{rf}}=0$~mT, the agreement is quantitative up to a certain value ($h_{\text{rf}}=1$~mT), after which there is no more agreement. In fact, the analysis of the spectrum obtained from micromagnetic simulations beyond this value shows that it is not SP anymore, but multi-peaks continuous (MP-C). 

This mismatch is investigated further with the reduced NMM. First, eq.\ref{eq:NMM_RWA} is linearized around a \textbf{P}-mode solution calculated by solving the \textbf{P}-mode equation with the continuation method, where the rf field is slowly increased. Then, for each $h_{\text{rf}}$ in the range $0-4$~mT, the eigenvalues of the linear dynamical system obtained are computed. The critical field for the occurrence of the instability process is defined as the lowest $h_{\text{rf}}$ where the real part of one of the eigenvalues becomes positive. This procedure is visualized in Suppl. Fig.\ref{fig:P-modes}(e)-(f). It corresponds to a generalization of the second-order Suhl's instability in the framework of normal modes. One main difference with the Suhl's approach is that the \textbf{P}-mode equation permits to consider large-angle steady oscillations before the instability.

\begin{figure}
	\centering
        \includegraphics[width=6cm]{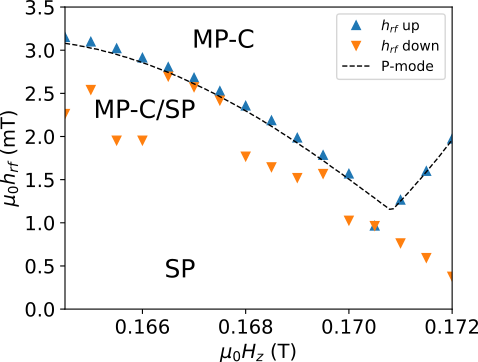}
	\caption{Phase diagram in the control plane ($H_z$, $h_{\text{rf}}$) calculated using the NMM including five modes (1, 6, 15, 30, 51). Instability curves with upwards (downwards) triangles have been obtained by increasing (decreasing) slowly $h_{\text{rf}}$ from zero (maximum) amplitude. The dashed curve is the instability threshold deduced from \textbf{P}-mode analysis (see Suppl. Fig.\ref{fig:P-modes}(f)).}
	\label{fig:NMMbistab}
\end{figure}

The instability curve defined in this way looks qualitatively similar but not in quantitative agreement with the curve found from micromagnetic analysis shown in Fig.\ref{fig:magico}(g) (no rate). So we investigated the dynamics with continuous time simulations: $H_z$ being fixed, $h_{\text{rf}}$ is first slowly increased to the point where the SP solution becomes an MP-C solution, and then decreased until the SP solution is recovered. This continuous time analysis has been done both with the micromagnetic code MaGICo and with the reduced NMM. In Fig.2(g) of the main text, the results obtained with MaGICo are shown, while those obtained with the reduced NMM are shown in Suppl. Fig.\ref{fig:NMMbistab}. They reveal several interesting facts. First, the instability curve found from micromagnetic simulations where the equilibrium was taken as initial condition is very close to the curve corresponding to the transition MP-C$\rightarrow$SP computed from continuous time micromagnetics. Second, the curve corresponding to the transition SP$\rightarrow$MP-C, that was a prediction of the \textbf{P}-mode analysis, has also been found with micromagnetic simulations. Third, the SP$\rightarrow$MP-C curve computed by the reduced NMM reproduces well that computed by micromagnetic simulations. Finally, the reduced NMM captures the coexistence between the SP and MP-C solutions. The region of coexistence between these two solutions differs from the one obtained from full micromagnetics because the NMM with only 5 modes describes the dynamics of the MP-C solution only qualitatively. 


\section{Experimental evidence of bistable dynamics}

\begin{figure}
	\centering
        \includegraphics[width=12cm]{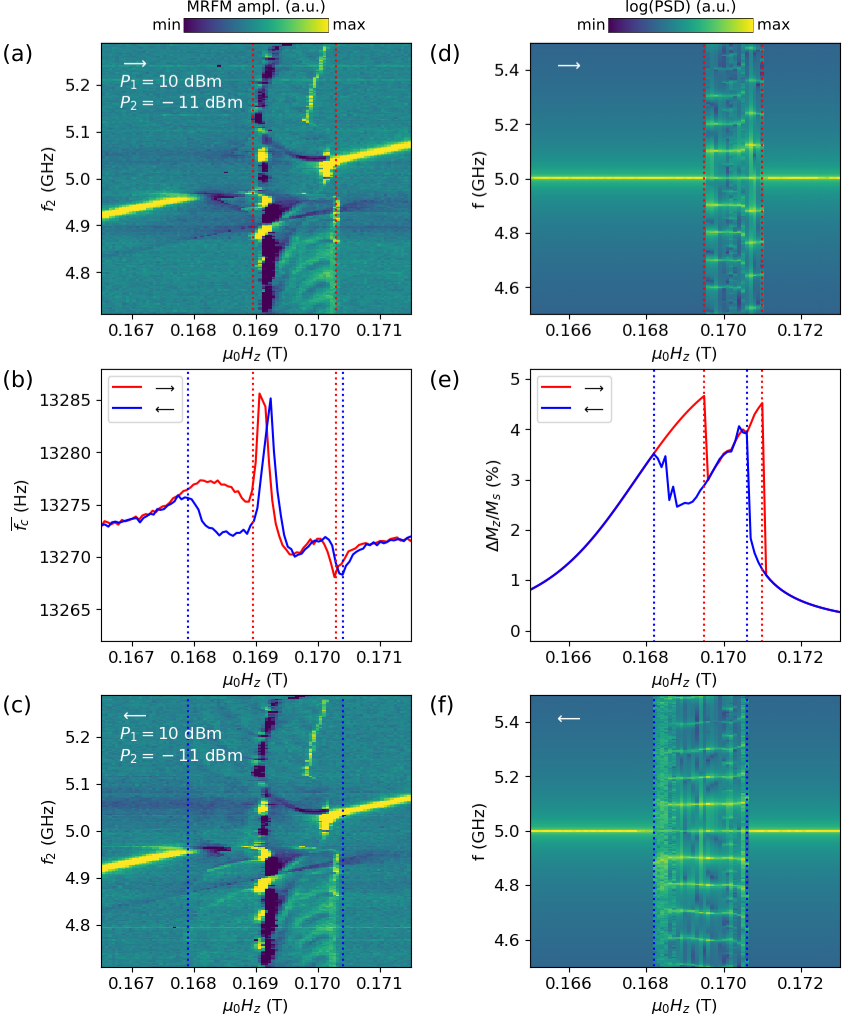}
	\caption{Experimental evidence of bistable dynamics. (a) and (c) Two-tone MRFM spectroscopy as a function of the upwards and downwards swept field $H_z$, respectively, and the frequency $f_2$ of the low power rf excitation (pulse modulated at $-11$~dBm). The power of the continuous rf excitation at 5~GHz is fixed to $10$~dBm. (b) Simultaneously measured MRFM cantilever frequency vs. $H_z$. (e) Longitudinal reduction of magnetization vs. $H_z$ extracted from micromagnetic simulations ($\mu_0h_{\text{rf}}=1$~mT at 5~GHz). (d) and (f) Corresponding spectral analysis as a function of the upwards and downwards swept $H_z$, respectively. Red and blue dotted vertical lines indicate limits of the instability region on all panels of the figure.} 
	\label{fig:bistabsup}
\end{figure}

To further demonstrate the bistable dynamics driven by nonlinear FMR, we provide additional experiments to the ones presented in Fig.5 of the main text. Suppl. Fig.\ref{fig:bistabsup} presents two-tone MRFM spectroscopy measurements performed on a BiYIG nanodisk of diameter 500~nm and their comparison to micromagnetic simulations. The evidence of bistability is achieved by sweeping up and down the applied magnetic field $H_z$ under continuous rf excitation at $f_1=5$~GHz and $P_1=10$~dBm, and by supplying a second rf tone of low power $P_2=-11$~dBm and varying frequency $f_2$ to probe the spectrum of the dynamics driven by the main rf tone. To better evidence a possible hysteretic behavior, it is important to sweep $H_z$ at constant $f_2$ rather than the opposite (which was the case in the two-tone MRFM spectroscopy maps shown in Fig.4 of the main text). 

Suppl. Fig.\ref{fig:bistabsup}(a) and (c) display the two-tone spectroscopy maps acquired by sweeping $H_z$ upwards and downwards, respectively. The comparison of these two panels demonstrates some hysteresis. For the upward sweep, a multi-peaks spectrum is measured from 0.169~T to 0.1703~T, meaning that in this region, the instability threshold is reached. In contrast, for the downward sweep, the multi-peaks dynamical state is measured from 0.1703~T all the way down to 0.168~T. In other words, for $0.168 < \mu_0H_z < 0.169$~T, the self-modulation instability is only observed upon decreasing the applied field. Additional evidence of this bistable dynamics is provided by monitoring the frequency $f_c$ of the MRFM cantilever using a phase-locked loop while sweeping up and down the field, which is done simultaneously to the two-tone MRFM spectroscopy, and presented in Suppl. Fig.\ref{fig:bistabsup}(b). One observes that the upward and downward traces of the $f_c$ signal do not superpose for $0.168 < \mu_0H_z < 0.169$~T, an indication that the static change of longitudinal magnetization induced by the continuous rf field is hysteretic in this region \cite{li19c}. We note that the specific shape of the $f_c$ signal in Fig.\ref{fig:bistabsup}(b) is due to the cantilever oscillations above the disk \cite{ngouagniayemeli22}.

Finally, we performed full micromagnetic simulations which reproduce these experiments by slowly sweeping $H_z$ up and down under continuous rf excitation. The longitudinal reduction of magnetization is shown in Suppl. Fig.\ref{fig:bistabsup}(e) for upward and downward sweeps, while the corresponding power spectra are displayed in panels (d) and (f), respectively. Similarly to the MRFM spectroscopy data, a clear bistable dynamics is observed for $0.1682 < \mu_0H_z < 0.1695$~T between the SP and MP-C states, respectively achieved in this region for increasing and decreasing $H_z$. We also note the appearance of a second narrower hysteretic region for $0.1706 < \mu_0H_z < 0.171$~T where these two dynamic states can be excited depending on the field sweep direction.

\clearpage
\newpage
\section*{References}

\let\oldaddcontentsline\addcontentsline
\renewcommand{\addcontentsline}[3]{}
\bibliography{NL-FMR}
\let\addcontentsline\oldaddcontentsline